# An automated system for strain engineering and straintronics of 2D materials

Onur Çakıroğlu[1], Joshua O. Island[2], Yong Xie[1,3], Riccardo Frisenda[4] and Andres Castellanos-Gomez[1*]

[1] *Materials Science Factory, Instituto de Ciencia de Materiales de Madrid (ICMM-CSIC), Madrid, E-28049, Spain*

[2] *Department of Physics and Astronomy, University of Nevada Las Vegas, Las Vegas, NV, 89154, USA*

[3] *School of Advanced Materials and Nanotechnology, Xidian University, Xi'an, 710071, China*

[4] *Department of Physics, Sapienza University, Roma, 00185, Italy*

*Correspondence: andres.castellanos@csic.es

This work presents an automated three-point bending apparatus that can be used to study strain engineering and straintronics in two-dimensional materials. We benchmark the system by reporting reproducible strain tuned micro-reflectance, Raman, and photoluminescence spectra for monolayer molybdenum disulfide ($MoS_2$). These results are in good agreement with reported literature using conventional bending apparatus. We further utilize the system to automate strain investigations of straintronic devices by measuring the piezoresistive effect and the strain effect on photoresponse in an $MoS_2$ electrical device. The details of the construction of the straightforward system are given and we anticipate it can be easily implemented for study of strain engineering and straintronics in a wide variety of 2D material systems.

Strain engineering is an efficient route to tune the electrical and optical properties of two-dimensional (2D) materials since their electronic band structures are highly sensitive to mechanical deformation of their lattices.[1–16] Moreover, van der waals (vdW) materials have strong mechanical properties and can withstand very high strain values, close to the breaking values predicted for ideal brittle materials, due to a lack of dangling bonds on their surfaces and low density of defects.[17–21] Thus, the development of 2D-based devices with output characteristics and performance that can be adjusted by means of externally applied strain (so called straintronic devices) are attracting great attention from the materials science and electronic devices communities.[22,23]

In most strain engineering experiments reported in the literature, strain is induced by bending flexible substrates on which the flakes are placed.[24–27] This method yields homogeneous uniaxial strain and it is very straightforward to implement.[28] However, in many of the experimental setups employed in the reported works, the flexible substrates are bent with manual actuators thus suffering from low accuracy and reproducibility of the applied strain, and hampering the acquisition of large datasets over long periods of time.

In this work, we propose an automated strain setup based on a motorized three-point-bending apparatus. First, we provide technical details to build up the experimental setup compatible with optical and electrical characterization. Then, we show that results acquired in differential reflectance, photoluminescence measurements and Raman spectrum of single layer $MoS_2$ are comparable with literature to check its validity to be used in strain engineering experiments. Lastly, we demonstrate the capability of this automated system to measure two illustrative straintronic devices based on single-layer MoS$_2$: a piezoresistor and a strain-tunable photodetector. We believe that our automated straining setup can have a large impact in the community studying strain engineering of 2D materials as it can be easily reproduced by other groups allowing acquisition of larger and more precise datasets.

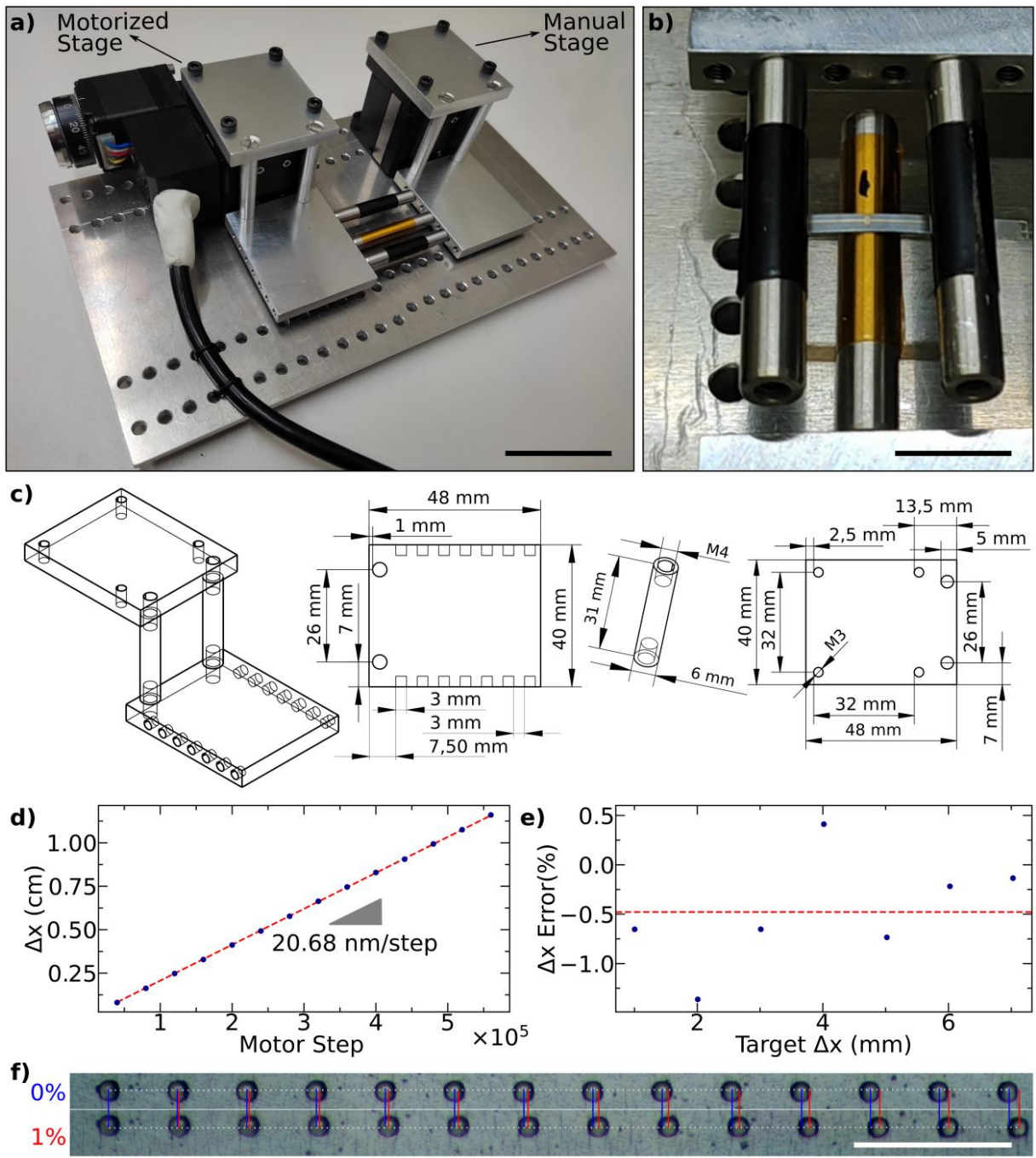

**Figure 1:** Motorized straining setup and its calibration. **a)** Picture of the complete setup. **b)** Picture of a flexible substrate with electrodes subjected to strain with the straining setup. **c)** Blueprints of the homebuilt parts needed to assemble the setup. **d)** Vertical displacement of the motorized stage ($\Delta x$) as a function of the number of motor steps. **e)** Error in the vertical displacement when moving the stage to a specific target height. **f)** Optical microscopy images of photoresist micropillars, patterned on the surface of a polycarbonate (PC) substrate, that allow the direct measurement of the strain applied to the flexible substrate. The blue and red lines mark the center of the pillars for the unstrained and strained substrate, respectively. Scale bars are 40 mm, 10 mm and 100 μm in **a**, **b** and **f** respectively.

The motorized straining setup consists of 3 main components; a motorized vertical translation stage (8MVT40-13-1, Standa), a manual stage (7VT40-13, Standa) and homebuilt parts shown in Figure 1 (a-d). Homebuilt parts are meant to mount the cylinders (MS1R/M, Thorlabs), that will be used as the three contact points of the three-point bending system, at a convenient position that allows inspection under a conventional optical microscope system. Two cylinders (loading pins) were attached to a homebuilt part, on the motorized stage, to bend the sample and one cylinder (support pin) was attached to the other homebuilt part, on the manual stage, to support the sample. The manual stage is convenient for adjusting the contact cylinder to hold the sample in place upon initial loading. The stages were mounted on an aluminum base to facilitate carrying the setup easily. The manual stage is used to secure the substrate between the three contact points when loading a new sample and it remains static during the experiment while the motorized stage is moved by a computer connected to a motor driver (8SMC5-USB-B8-1, Standa). Supporting Information Figure S1 shows a version of the automated straining setup featuring homebuilt parts printed with a 3D printer (Elegoo Mars 2 printer). We address the reader to the Supporting Information for the source files to print the required homebuilt parts.

The motorized stage has been calibrated to determine the displacement associated with each motor step. To do so, the motor is set to move a given number of motor steps and the actual displacement of the stage is measured by quantitative analysis of optical images and by direct measurements with a caliper. This process is repeated for different number of motor steps providing a relationship between the number of motor steps and the actual displacement of the stage (see Figure 1(d)). The slope of the displacement vs. motor step gives the step size of the stage as 20.68 nm. The difference between the target displacement, set for the motorized stage with the computer, and the measured displacement provides a value for the experimental error. As shown in Figure 1(e) the stage displacement is highly accurate: a typical target displacement of 1 mm will be off by only 5 μm.

The amount of strain ($\varepsilon$) induced with the displacement of the motorized stage in the three-point bending setup can be calculated with the formula[1]

$$\varepsilon = \frac{6Dt}{L^2} \qquad Eq.\ 1$$

where $D$, $t$ and $L$ are the displacement of the contact points, the thickness of substrate and the distance between the outer two contact points. Assuming the typical thickness of our flexible substrates (250 μm) and a contact point distance of 18 mm, the motorized stage allows for modifying the strain in steps of only ~$10^{-6}$%. Complementary to the use of Eq. 1, strain can also be directly measured using a flexible sample with features on its surface separated by a given distance and measuring the distance between them before and after applying strain. Figure 1(f) shows optical microscopy images of an array of photoresist micro-pillars, patterned on the surface of a polycarbonate (PC) substrate by optical lithography, before (top image) and after (bottom image) applying a strain of 1%. These are patterned on the same substrate used for all the straining experiments reported here. The distance between the micropillars at zero strain ($l_0$) and at a given applied strain ($l$) provides the experimental values of strain by the formula $\varepsilon = (l-l_0)/l_0$.[28] To find this distance, the center of the micropillars are marked with vertical lines, the distance between these lines are measured to determine $l$ and $l_0$. We found that the experimental strain values match the strain calculated with Eq. 1 within the experimental uncertainty (see Supporting Information Figure S2).

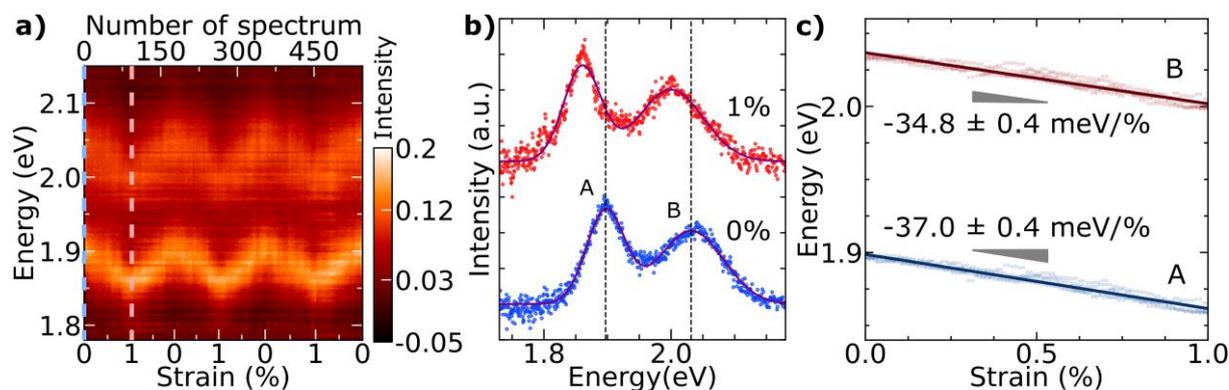

**Figure 2:** Strain tunable differential reflectance in monolayer $MoS_2$. **a)** Evolution of the differential reflectance of $MoS_2$ in 3 strain load/unload cycles. The colormap shows the intensity of the differential reflectance in the color axis, the horizontal axis the uniaxial strain (in %, upper axis shows the number of the spectrum) and the vertical axis the energy. The measurement is composed of 546 differential reflectance spectra, acquired continuously. **b)** Comparison of two differential reflectance spectra at different strain values. The 0% and 1% strain spectra correspond to vertical line-cuts in **a)** highlighted with the dashed light blue and red vertical lines respectively. Dash lines in **b)** show positions of A and B exciton peaks for unstrained flake. **c)** A and B exciton peak energies as a function of tensile strain. The datasets (composed of 546 individual measurements) were fit to a linear trend to extract the gauge factor.

We have tested the performance of the motorized straining setup to tune the optical properties of 2D materials by continuously acquiring differential reflectance spectra of a single-layer $MoS_2$ flake while the motorized setup cycles the strain from 0% to 1% uniaxial tension three times. We have intentionally limited the maximum strain to 1 % to prevent slippage that can occur at strains higher than 0.8 %.[28] It should be noted though that the apparatus is capable of reaching much higher strains. See the Supporting Information Figure S3 for a straining experiment reaching 2% uniaxial tension where the flake clearly displays slippage for strains larger than ~1.5%. The $MoS_2$ flake is transferred to the flexible PC substrate using a deterministic transfer method.[29] The sample is then loaded into the apparatus by using the manual stage to lightly pin the substrate in place. Note that the focus had to be slightly re-adjusted during the straining cycles and a refocus of less than ~75 μm was needed during the cycles. The motorized system allows one to acquire large datasets in a short time: more than 500 spectra in ~8 minutes. This can be very useful when compared to manual straining setups that are capable of collecting 5-10 datapoints in roughly 30 mins time. Figure 2(a) shows a colormap created with 546 differential reflectance spectra acquired while the motorized setup applies 3 cycles of loading/unloading uniaxial strain (from 0% to 1%, see lower horizontal axis). The brightest zones around 1.90 and 2.03 in Figure 2(a) are a result of the A and B excitons, which originate from direct transition between valence and conduction bands and transition between lower lying valence and conduction bands of single-layer $MoS_2$ at the K point of the Brillouin zone.[30–32] The resonant energy of the A and B excitons decreases with higher strain values and increases back to unstrained values at 0%. Figure 2(b) compares two spectra acquired at 0% and 1% uniaxial strain values taken at the location of the vertical dashed lines in Figure 2(a). The experimental data points (dots) are fit to two Gaussians to accurately determine the energy of the A and B excitons. Figure 2(c) illustrates how both the A and B exciton energies shift towards lower values upon increase of uniaxial strain in a linear fashion. The experimental data can be fit to a linear trend to determine the differential reflectance gauge factor, i.e. the shift of the A and B exciton energies in the differential reflectance spectra per % of uniaxial tensile strain. The gauge factor values were found to be -34.8 ± 0.4 meV/% and -37.0 ± 0.4 meV/% for A and B excitons respectively. These values are in good agreement with literature.[28,33,34] We refer the reader to the Supporting Information (Figure S4) for a comparison direct comparison between data acquired on a manual strain set up and a manual one, highlighting the improvements in data collection and reduction in statistical error.

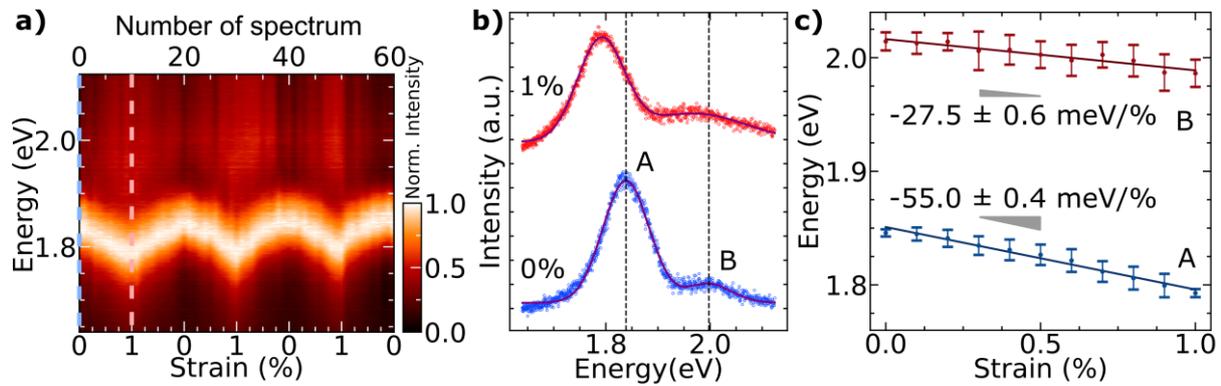

**Figure 3:** Strain tunable photoluminescence (PL) measurements of monolayer MoS$_2$. **a)** Evolution of PL of MoS$_2$ in 3 strain load/unload cycles. The colormap shows the intensity of the PL, the horizontal axis the uniaxial strain (in %, upper axis shows the number of measurements) and the vertical axis the energy. The measurement is composed of 60 photoluminescence spectra. **b)** Comparison of two PL at different strain values. The 0% and 1% strain spectra correspond to vertical line-cuts in **a)** highlighted with the dashed light blue and red vertical lines respectively. Dash lines in **b)** show positions of A and B excitons for unstrained flake. **c)** PL peak energies as a function of tensile strain. The datasets (composed of 60 individual measurements) were fitted to a linear trend to extract the gauge factor.

We have also tested the operation of the motorized stage in photoluminescence measurements. Photoluminescence spectra were acquired on another single-layer MoS$_2$ sample while the motorized setup applied three straining/unloading cycles. Note that, unlike our homebuilt differential reflectance system, the photoluminescence setup (MonoVista CRS+, Spectroscopy & Imaging GmbH, excitation laser wavelength of 532 nm, laser spot ~ 800 nm) does not allow for sample inspection during the spectra collection making the in-situ slight refocus operation not feasible. We thus performed the cycles in a step-wise fashion, checking the focus at each step and slightly re-adjusting when needed. Each cycle consisted of 10 steps increasing strain up to 1% and 10 steps decreasing strain down to 0%, collecting a total of 60 spectra in 220 minutes. Figure 3(a) shows a colormap consisting of 60 individual PL measurements acquired during 3 cycles of loading/unloading uniaxial strain. A and B excitons lead to higher intensity zones which reveal that the effect of the strain is reversible. The red shift is clearly observed comparing two spectra in Figure 3(b), taken at the location of the vertical dashed lines in Figure 3(a), while the peak positions of excitons are extracted by fitting the experimental data points (dots) to two Gaussian (solid lines). Applying the same procedure to 60 individual PL spectra helps to investigate redshift rates so that the rates (or gauge factor) in Figure 3(c) can be found as -55.0 ± 0.4 meV/% and -27.5 ± 0.6 meV/% for A and B exciton peaks respectively, which are also in close agreement with Refs [25,28,33–38].

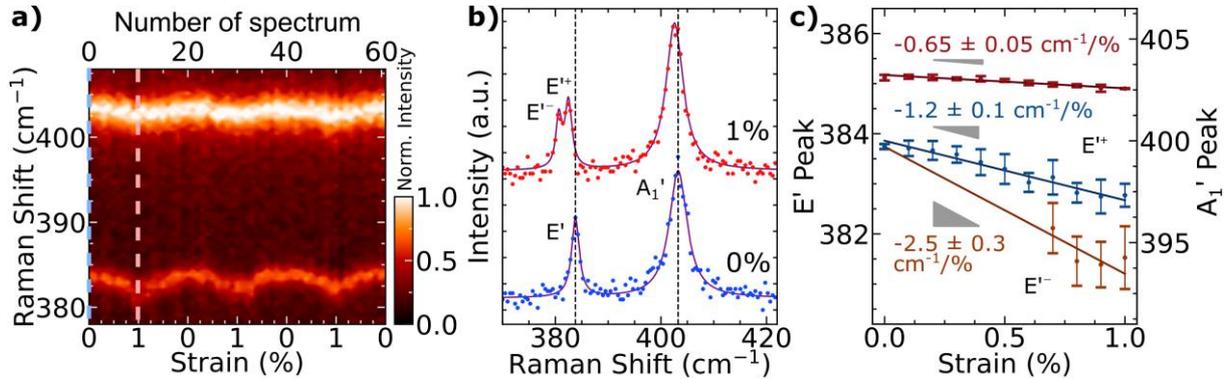

**Figure 4:** Strain tunable Raman spectrum of monolayer MoS$_2$. **a)** Evolution of Raman spectra of MoS$_2$ in 3 straining cycles. The colormap shows the normalized intensity of the Raman spectra, the horizontal axis is the uniaxial strain (in %, upper axis shows the number of measurements) and the vertical axis the Raman shift. The measurement is composed of 60 Raman spectra. **b)** Comparison of two Raman spectra at different strain values. The 0% and 1% strain spectra correspond to vertical line-cuts in **a)** highlighted with the dashed light blue and red vertical lines respectively. Dash lines in **b)** show positions of E′ and A$_1$′ Raman modes of unstrained flake. **c)** Raman modes as a function of tensile strain. Note that the E′ mode splits into two peaks after 0.7% strain value. The datasets (composed of 60 individual measurements) were fitted to a linear trend to extract the gauge factor.

Raman spectroscopy is another versatile tool to study 2D materials and it is found to be very sensitive to strain.[39,40] We therefore used our motorized stage to demonstrate its operation to strain-tune the Raman spectra. We evaluate Raman spectra upon uniaxial straining by acquiring 60 spectra while strain was applied in loading/unloading cycles. Raman spectroscopy was performed in the same system that we used for the PL spectroscopy. Thus, the cycles were performed in a stepwise fashion again, checking the focus at each step and slightly re-adjusting the focus when needed. Figure 4(a) shows a colormap created from the spectra (shown in the horizontal upper axis) during three straining/unloading cycles (strain shown on the horizontal lower axis). The Raman spectra show two prominent Raman peaks at 384 cm$^{-1}$ (so called E′, which originates from in-plane vibration mode of phonons in MoS$_2$ crystals) and at 403 cm$^{-1}$ (so called A$_1$′, which originates from out of plane vibration mode of phonons in MoS$_2$ crystals).[41,42] Both Raman modes shift towards lower Raman shift values upon uniaxial strain.[43] Interestingly, the E′ Raman mode is a degenerate phonon mode and it splits to two modes (E′$^-$ and E′$^+$) at uniaxial strain values higher than 0.7%.[43] Figure 4(b) compares two experimental Raman spectra acquired at 0% and 1% strain values. The dots are experimental data points and solid lines are a Lorentzian fit to accurately extract the center, amplitude and full-width-at-half-maximum of peaks. One can clearly see the shift towards lower Raman shift values of the peaks and distinguish the splitting of the E′ Raman modes at 1% strain. The shifts of the Raman modes upon strain, or the Raman modes gauge factors, can be extracted by fitting the peak locations, found by Lorentzian fit, to linear trends. We found values of -0.65 ± 0.05 cm$^{-1}$/%, -1.19 ± 0.11 cm$^{-1}$/% and -2.5 ± 0.3 cm$^{-1}$/% for the A′$_1$, E′$^+$ and E′$^-$ peaks, respectively. (see Figure 4(c))

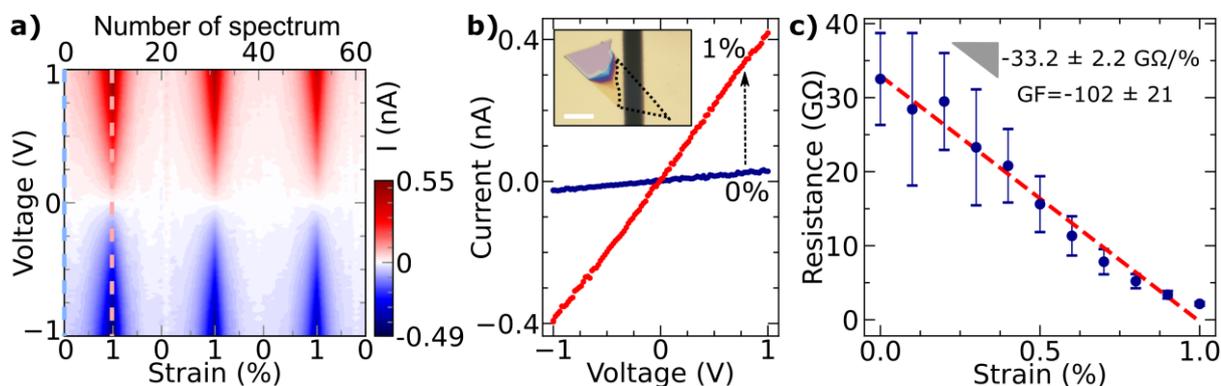

**Figure 5:** Piezoresistive effect in monolayer MoS$_2$. **a)** Evolution of current vs. voltage characteristics of a MoS$_2$ device during 3 strain load/unload cycles. The colormap shows the value of the current, the horizontal axis is the uniaxial strain (in %, upper axis shows the number of measurements) and the vertical axis the voltage. The measurement is composed of 63 current-voltage curves. **b)** Comparison of two current-voltage curves at different strain values. The 0% and 1% strain current-voltage curves correspond to vertical line-cuts in **a)** highlighted with the dashed light blue and red vertical lines respectively. Inset shows the optical image of the MoS$_2$ device on a PC substrate with gold contacts. The dashed line shows the monolayer part of the flake. Scale bar is 10 µm. **c)** Resistance as a function of tensile strain. The datasets (composed of 63 individual measurements) were fit to a linear trend and divided by the resistance value at 0% strain to extract the gauge factor.

Besides the changes observed in photoluminescence and reflectance, the change in bandgap upon strain should give rise as well to a change in the electrical properties of MoS$_2$. In the following we will discuss how the stability and accuracy of the motorized stage can be an important asset to study the performance of strain-tuneable electronic devices based on 2D materials, so called straintronic devices. A single-layer MoS$_2$ flake was deterministically transferred onto a PC substrate bridging two pre-patterned drain-source contacts. Only 50 nm thick Au electrodes were evaporated on the PC substrate by electron-beam evaporation using a metal shadow mask from Ossila® with a part number E292 before the transfer (see the inset in Figure 5(b)). We have measured current vs. voltage (*IV*s hereafter) characteristics of the single layer device while the motorized setup cycles simultaneously the strain from 0% to 1% three times to probe the effect of strain on electrical properties of the MoS$_2$ flake. Figure 5(a) shows a colormap consisting of 63 individual *IV* measurements during 3 cycles of loading/unloading strain. The change in the contrast of the colormap indicates that current increases (decreases) by increasing (decreasing) the strain. Figure 5(b) shows two *IV*s acquired at 0% and 1% strain to facilitate a quantitative comparison. These two IVs are taken at the location of the vertical dashed lines in Figure 5(a). The reversible change in resistance with strain is defined as the piezoresistivity effect, commonly used in strain gauge sensors. The effect is caused by a change in the energy band of the materials with strain which changes the probability of electrons to transition to the conduction band.[44] The resistance of the flake in each *IV* measurement can be calculated by its slope to characterize the piezoresistivity effect in the MoS$_2$ flake. Figure 5(c) summarizes the obtained resistance values as a function of the applied strain. By fitting the resistance vs. strain values, obtained from the 63 *IV*s, to a linear trend one can determine the piezoresistive gauge factor as $\Delta R/(R_0 \cdot \Delta \varepsilon)$ where $\Delta R$ is the change in resistance (extracted by the linear fit), $R_0$ is resistance of the unstrained flake and $\Delta \varepsilon$ is the change in strain respectively.[44] The gauge factor was found as -102 ± 21, of the same order as that reported in the literature.[44] In Ref [44], however, the authors applied strain with a sharp tip resulting in an inhomogeneous strain distribution combining biaxial and uniaxial strain on the suspended MoS$_2$ flake. In our experiment, the motorized stage leads to a uniform uniaxial strain distribution over the whole device and it is thus more straightforward to extract the piezoresistive gauge factor.

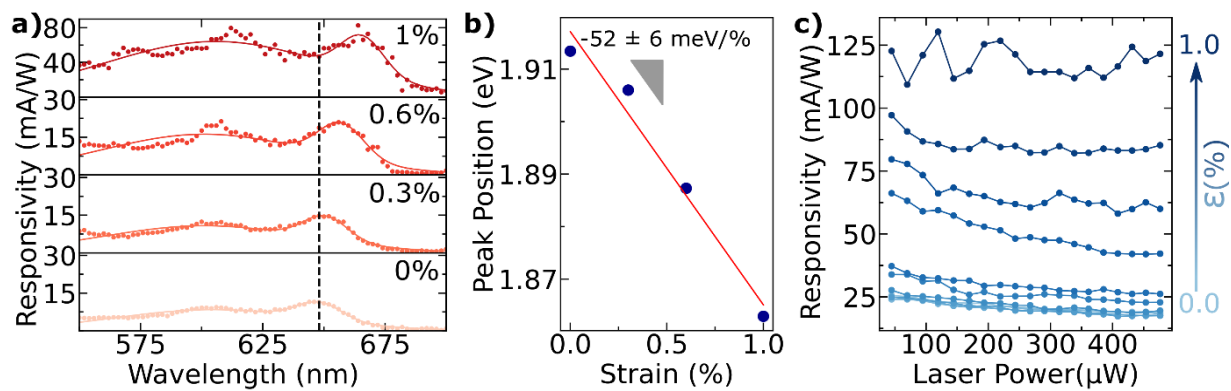

**Figure 6: Strain tunable photoresponsivity of monolayer MoS$_2$.** a) Comparison of wavelength dependent responsivity at different strain values. The vertical dashed line shows position of A exciton of unstrained flake. Dots show measured data points and solid lines show Gaussian fitted curves. b) A exciton energy as a function of tensile strain. The datasets (obtained from a) were fitted to a linear trend to extract the gauge factor. c) Comparison of laser power dependent responsivity at different strain values. Wavelength of the laser is 660 nm.

Finally, we use the motorized straining setup to tune the photoresponse of a single-layer MoS$_2$ based photodetector device. The device is the same one used for the electrical characterization upon strain shown in Figure 5. Because of the strain tunable band gap of MoS$_2$, one can tune the spectral response of a MoS$_2$-based photodetector by applying an external mechanical deformation.[22] Photocurrent is acquired by applying a bias voltage across the device and measuring the current both in dark conditions and under illumination. The responsivity value ($R$) can be then extracted by the formula $R = I_{ph}/P_{ch}$ where $I_{ph}$ is photocurrent, $P_{ch}$ is the light power reaching 1L MoS$_2$ channel.[45] The spectral response of the photodetector device has been measured at 4 different uniaxial strains in a wavelength range of 550 nm to 700 nm (in 2 nm increments) as shown in Figure 6(a). We used a xenon lamp with a monochromator (Bentham TLS120Xe), projecting a light spot of 1.6 mm in diameter over the device, to select the illumination wavelength. The photoresponsivity of 1L-MoS$_2$ has two prominent peaks resulting from the enhanced optical absorption due to the A and B excitons.[30–32] The experimental data points are fit to two Gaussians to determine their energy values. Similar to changes in the reflectance and PL experiments, the A exciton energy shifts towards higher wavelengths (or lower energy values) upon increase of uniaxial strain. Figure 6 (b) shows that the gauge factor is -52.0 ± 6.1 meV/% by extracting the A exciton energies from the spectral response. Figure 6 (a) also shows how uniaxial strain seems to have an effect on the magnitude of the responsivity: the higher the uniaxial strain the more sensitive the photodetector becomes. We plot the amplitude of the A exciton peak as a function of strain in Figure S5 in the supporting information. We further study the effect of uniaxial strain in the responsivity by performing a comprehensive responsivity vs light power measurement at 11 different strain values (0% to 1%) for a fixed wavelength of 660 nm using an LED (Thorlabs M660FP1). Figure 6 (c) shows that the responsivity increases upon increasing the uniaxial strain and there is a slight decrease with increasing light power due saturation of trap states that contribute to the photogating effect in the photocurrent generation.[22,46] We attribute this increase of the responsivity upon increasing tensile strain to the dramatic electrical resistance strain-induced reduction which leads to shorter drift times of the charge carriers in the channel and thus higher photogating gain.[47–49] In the Supporting Information Figure S6 we show the response time of the MoS$_2$ photodetector for 0% and 1% uniaxial strain where one can infer an almost strain independent response time of the detector of ~150 ms.

We developed an automated strain setup for higher accuracy and reproducibility of applied strain and to allow automated acquisition of large datasets. We provide all technical details of the setup. We performed differential reflectance, photoluminescence and Raman spectroscopy on single-layer MoS$_2$ to test our straining setup and to show its applicability in strain engineering experiments for 2D materials. We further demonstrate the suitability of the motorized straining setup to characterize straintronic devices. We studied two proof-of-concept straintronic devices based on single-layer $MoS_2$: a piezoresistor and a photodetector with strain-tunable spectral bandwidth and responsivity. We anticipate that this experimental setup could be easily reproduced in other labs interested in strain engineering of 2D materials which could benefit from the improved accuracy and reproducibility of this automated setup and from the capability of acquiring datasets not possible with manually actuated straining setups.

NOTE: During the elaboration of this manuscript we became aware of the work of Pop's and co-workers [50] reporting a piezoresistive gauge factor of -200 ± 45 /% for chemical vapour deposition grown $MoS_2$ single-layer.

**Funding**

European Research Council (ERC) (755655)
European Union's Horizon 2020 research and innovation program (956813)
EU FLAG-ERA (JTC-2019-009)
Comunidad de Madrid (Y2020/NMT-6661)
Spanish Ministry of Science and Innovation (PID2020-118078RB-I00)

**Acknowledgements**

We acknowledge funding from the European Research Council (ERC) through the project 2D-TOPSENSE (GA 755655), European Union's Horizon 2020 research and innovation program under the grant agreement 956813 (2Exciting), EU FLAG-ERA through the project To2Dox (JTC-2019-009), Comunidad de Madrid through the project CAIRO-CM project (Y2020/NMT-6661) and Spanish Ministry of Science and Innovation through the projects PID2020-118078RB-I00.

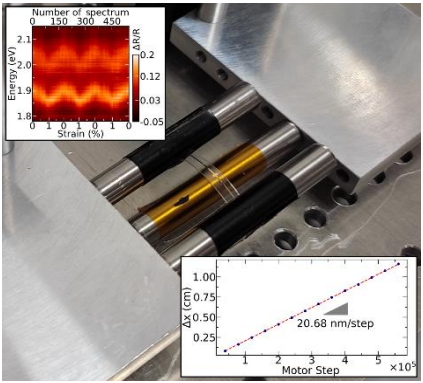

An automated three-point bending apparatus to study strain engineering and straintronics in two-dimensional materials is presented. The system allows one to perform several straining cycles in an unattended way with improved reproducibility and stability. The system is benchmarked with several strain engineering experiments in single-layer $MoS_2$.

# Supporting Information:

## An automated system for strain engineering and straintronics of 2D materials

Onur Çakıroğlu[1], Joshua O. Island[2], Yong Xie[1,3], Riccardo Frisenda[4] and Andres Castellanos-Gomez[1*]

[1] *Materials Science Factory, Instituto de Ciencia de Materiales de Madrid (ICMM-CSIC), Madrid, E-28049, Spain*

[2] *Department of Physics and Astronomy, University of Nevada Las Vegas, Las Vegas, NV, 89154, USA*

[3] *School of Advanced Materials and Nanotechnology, Xidian University, Xi'an, 710071, China*

[4] *Department of Physics, Sapienza University, Roma, Italy*
*Correspondence: andres.castellanos@csic.es

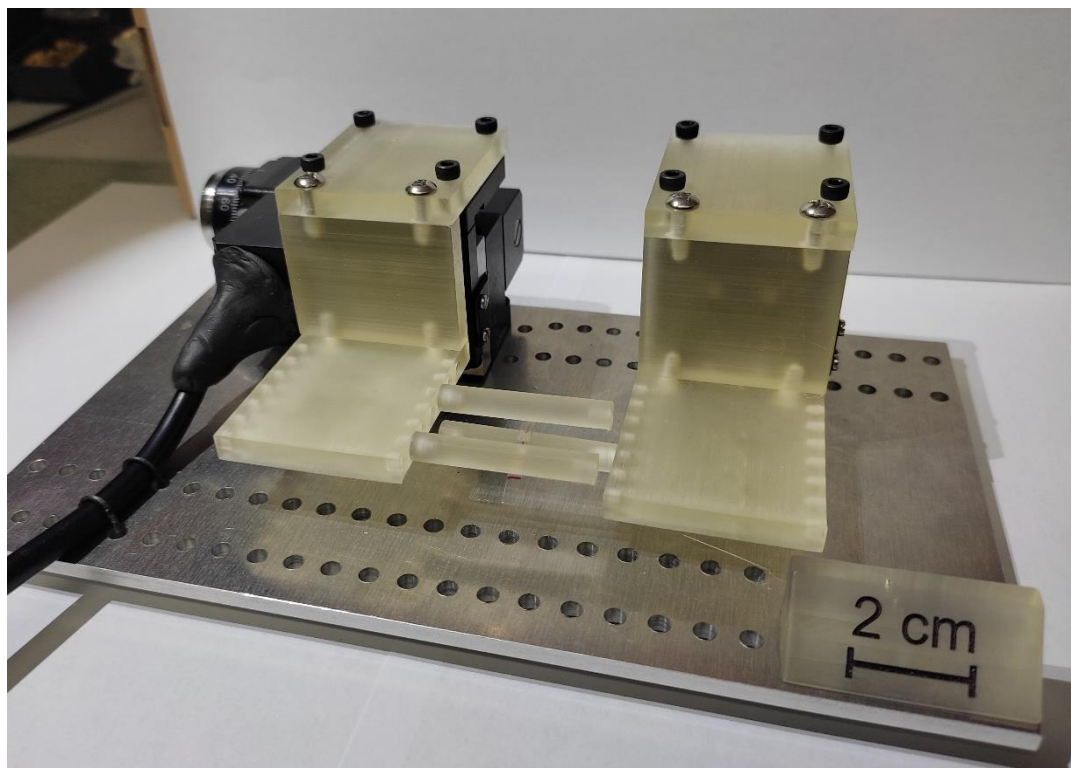

**Figure S1:** Picture of the automated straining setup with homebuilt parts printed with a 3D printer ELEGOO Mars 2. Translucent resist (ELEGOO LCD UV 405nm 3D) was used and an ANYCUBIC Wash & Cure Machine 2.0 was using for the improved curing of the parts. In the Supporting Information we attach the '*.FCSTD' files needed to print these 3D printed parts to allow an easy replication of this setup.

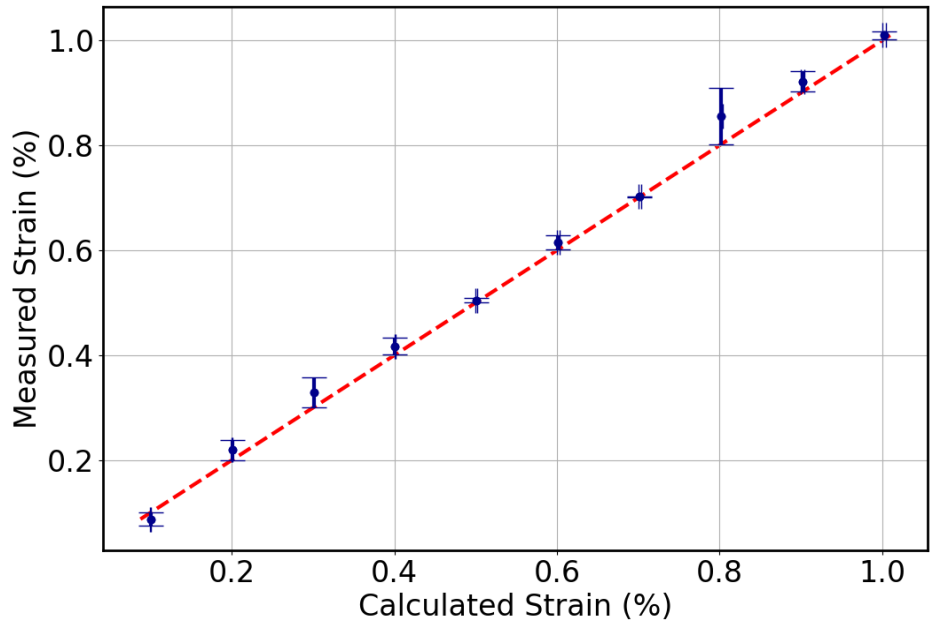

**Figure S2:** Verification of Equation 1 in the main text. The calculated strain, using Equation 1, matches well the strain from experimental measurements extracted from the distance of the pillars as a function of the deflection.

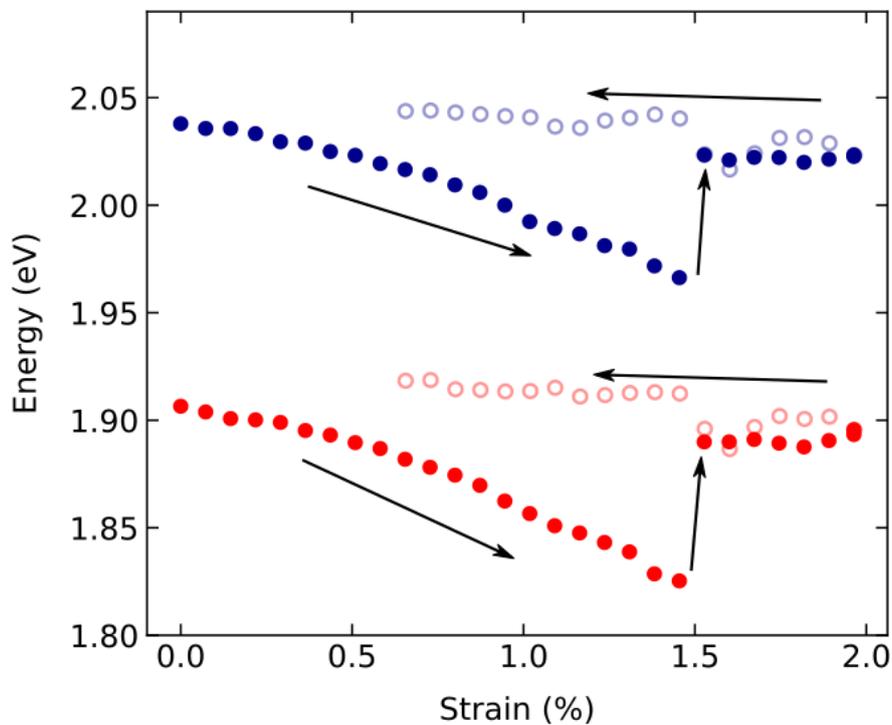

**Figure S3:** Example of a straining experiment using the motorized 3-point bending setup reaching strains above 1%. At ~1.5% strain there is a sudden jump of the energy of the excitons, attributed to relaxation of the accumulated strain due to slippage and since that point the strain is not properly transferred to the flake. This can be seen by the flat Energy vs.

strain trend of the datapoints after the strain relaxation. The empty light colored points have been acquired while releasing the strain.

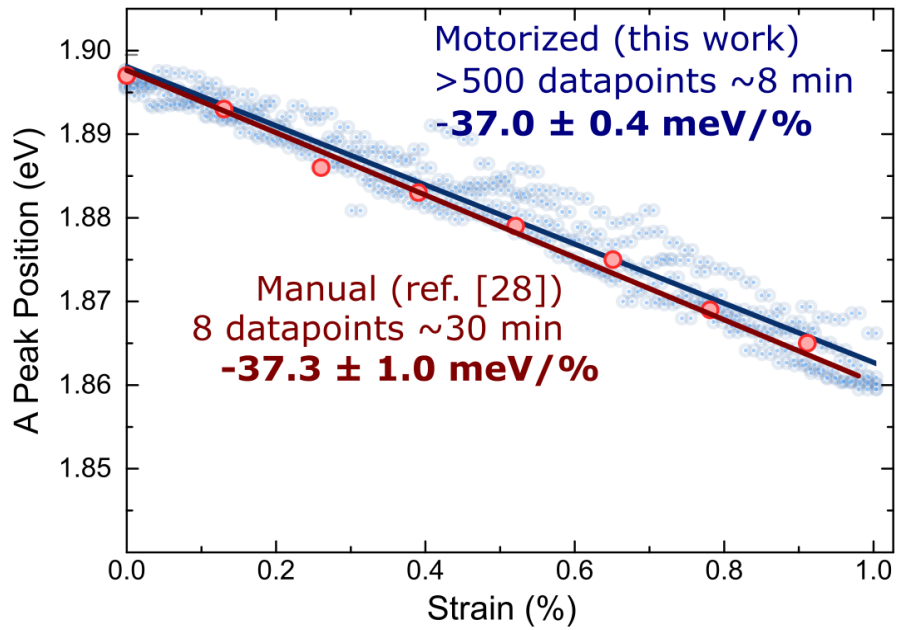

**Figure S4:** Comparison between two datasets of the gauge factor of the A exciton of monolayer MoS2, one obtained from a manual setup (see Ref. [28] of the main text) and one obtained with our automatic straining setup. The automatic setup allows the collection of a large amount of data over a shorter amount of time. Due to the larger dataset, there is also a lower statistical error in the determination of the gauge factor.

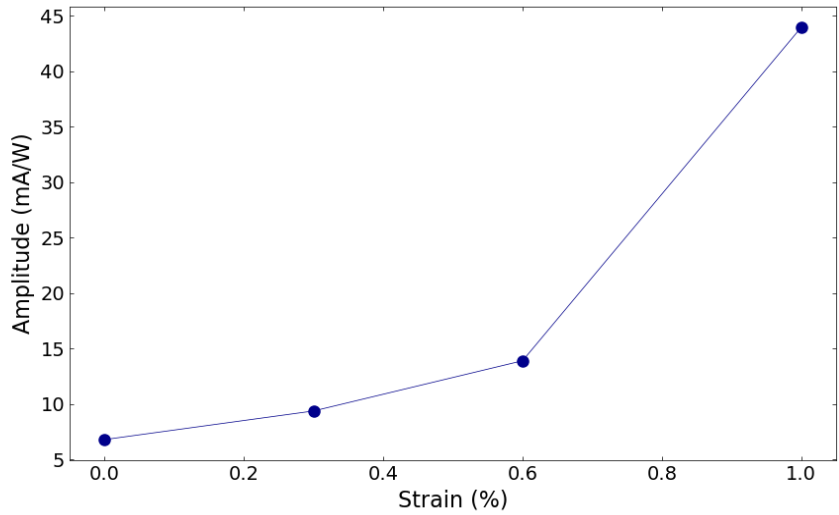

**Figure S5:** Strain dependent A exciton amplitude, extracted from Figure 6a of the main text.

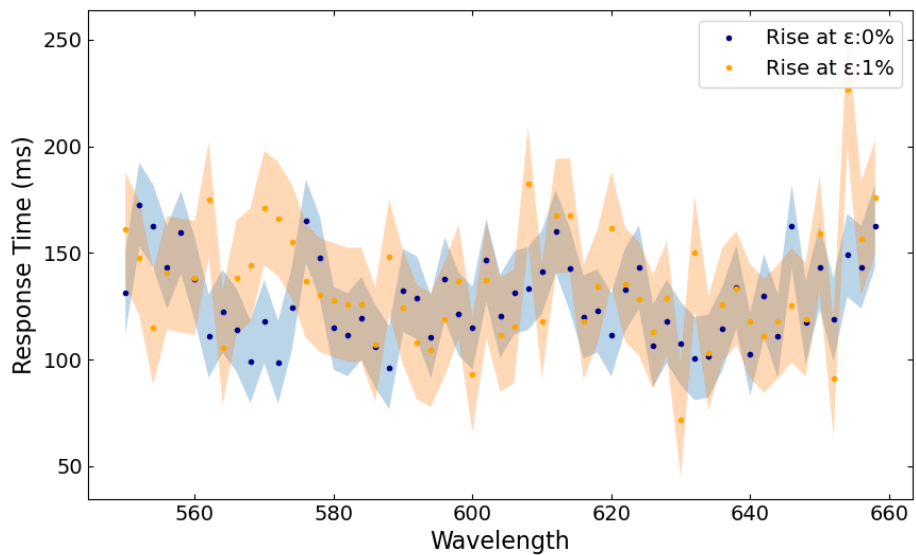

**Figure S6:** Response time of the MoS$_2$ straintronic photodetector for different strain values (0% and 1%) and for different wavelengths of light excitation.